\documentclass[journal]{IEEEtran}
\IEEEoverridecommandlockouts

\ifCLASSINFOpdf
   \usepackage[pdftex]{graphicx}
  \DeclareGraphicsExtensions{.pdf,.jpeg,.png}
\else
\fi
\ifCLASSOPTIONcompsoc
  \usepackage[caption=false,font=normalsize,labelfont=sf,textfont=sf]{subfig}
\else
  \usepackage[caption=false,font=footnotesize]{subfig}
\fi

\usepackage{stfloats}

\usepackage{booktabs} 
\usepackage{ulem}
\usepackage{color}
\usepackage{amsmath}

\begin{document}
%
\title{Effects of a Coal Phase-Out on Market \\Dynamics: Results from a Simulation Model\\for Germany}

\author{\IEEEauthorblockN{
Ramiz Qussous\IEEEauthorrefmark{1}, 
Thomas K\"unzel\IEEEauthorrefmark{2} and 
Anke Weidlich\IEEEauthorrefmark{1}}\\
\IEEEauthorblockA{\IEEEauthorrefmark{1}Department of Sustainable Systems Engineering (INATECH), University of Freiburg, 79110 Freiburg, Germany\\
Email: ramiz.qussous@inatech.uni-freiburg.de}\\
\IEEEauthorblockA{\IEEEauthorrefmark{2}Fichtner GmbH \& Co. KG, Stuttgart, Germany}



\thanks{
{\copyright} 2019 IEEE. Personal use of this material is permitted. Permission from IEEE must be obtained for all other uses, in any current or future media, including reprinting/republishing this material for advertising or promotional purposes, creating new collective works, for resale or redistribution to servers or lists, or reuse of any copyrighted component of this work in other works.}
}

\IEEEoverridecommandlockouts

\maketitle
\IEEEpubidadjcol
\pagestyle{plain}


\begin{abstract}
Modeling power market dynamics is increasingly challenging, as both spatial and temporal imbalances of demand and supply are becoming more pronounced with higher shares of variable renewable energy (VRE). Therefore, a high-resolution spatial and temporal distribution of both VRE and conventional generation, as well as the load, is necessary for modeling the effect of policy measures on the electricity market. Besides, technical constraints must be considered in detail, as they fundamentally influence market outcomes. This contribution presents the results and findings from a power market simulation model for the case of the German market zone under different scenarios phasing out coal-fired power generation. It is shown that the developed spatially resolved model can realistically reproduce market outcomes and is, therefore, suitable for the analysis of future scenarios with increased VRE integration and reduced conventional generation capacity.
\end{abstract}


%
\IEEEpeerreviewmaketitle

\section{Introduction}

In December 2014, the Climate Action Program 2020 was adopted by the German government, with the ambitious plan of reducing greenhouse gas emissions by 40\,\% compared with 1990 levels by 2020 \cite{BMU2018}. Although Germany has witnessed a successful expansion of renewable capacities, which provided it with about 157\,TWh of electricity in the year 2018, a large share of electricity generation still comes from power plants fired by fossil fuels, with coal-fired generation making up around 204\,TWh in 2018 \cite{BMU2018,ENTSOE.2019}. 
The decision on phasing-out nuclear power generation by 2022 makes the \smash{CO\raise-0.5ex\hbox{\scriptsize{2}}} reduction goal even more challenging to achieve. In order to close the apparent gap, an accelerated coal phase-out has been advised by the Commission on Growth, Structural Change, and Employment in her final report \cite{BMWI2019}, suggesting the end of the usage of coal for power generation by the year 2038 at the latest.

A phase-out of coal-fired generation will have significant effects on the energy market results and dynamics. One expected effect is a higher market share for flexible gas-fired power plants, partnering generation based on renewable energy \cite{Oei2019}. In this work, three different coal phase-out scenarios are modeled for the years 2022 and 2030. Their impact on power prices and plant deployment are studied with the help of a detailed electricity market simulation model.  

Since market liberalization, many detailed energy system models have been developed for facilitating power market analysis. According to \cite{Ventosa.2005}, these can be classified into the three categories of optimization, equilibrium, and simulation models. Among these, optimization and some simulation models have a high level of technological explicitness, as they consider the relevant technical constraints affecting generation units in detail, and they model individual generation units in a non-aggregated manner. For example, \cite{Ellersdorfer.2009} presents an optimization model applying game theory concepts. In the underlying optimization problem, a large number of technical parameters are taken into account through constraints, which restrict the flexibility of the generation units. \cite{Scholz.2012} develop a deterministic optimization model aiming at finding the cost-optimal design of the power supply system. \cite{Rosen2008} provide an example of a long-term comprehensive optimization model. The model is a technology-based energy and material flow model, which optimizes production and investment decisions within different energy systems. Some electricity market optimization models are also coupled with network grid models, as can be seen in \cite{Egerer2016Open}, or \cite{Koch.2015}. Such models are used to find cost-optimal solutions while considering network constraints. In \cite{Koch.2015}, control reserve requirement is simplified and taken into account via a base load to be covered all year around by thermal generation units. \cite{Egerer2016Open} focuses on spot markets and congestion management, but does not consider control reserve provisioning. 

Among the simulation models, agent-based models of electricity markets are very popular; \cite{weidlich2008critical} provide a survey of many of these approaches. Among the agent-based simulation models are some large-scale models of markets in different countries, such as PowerACE for Germany \cite{weidlich2008studying,SENSFUSS}, NEMSIM for Australia \cite{BattenGrozev}, AMES for New England (USA) \cite{Krishnamurth}, or EMCAS \cite{Conzelmann}, which has been developed in the United States and was applied to different national markets. 

While many of the mentioned models are highly accurate in representing individual plants with their technical constraints, they usually do not model the market interaction of plant operators in detail. This makes them less suitable for estimating price developments in future market configurations. Besides, the interrelations between different markets in the power sector, such as control reserve markets, or heat marketing by combined heat and power plants, are usually not modeled explicitly. In real markets, theses aspects are very relevant, so that they need to be taken into account, which is done in the model used for the analysis presented here.

The remainder of the paper is structured as follows: Section~\ref{sec:modelOverview} describes the proposed model, while Section~\ref{sec:scenarioOverview} reviews the scenarios and assumptions investigated in this contribution. Section~\ref{sec:results} discusses the model results, and Section~\ref{sec:conclude} finally concludes.

\section{The Model}\label{sec:modelOverview}

In the following, the modeling approach followed in this work is presented. Section~\ref{subsec:structure} gives an overview of the model structure, and Section~\ref{subsec:sequence} describes the simulation sequence.

\subsection{Overview of the Model Structure}\label{subsec:structure}

\begin{figure*}[!hb]
\centering
\subfloat{\includegraphics[width=\linewidth]{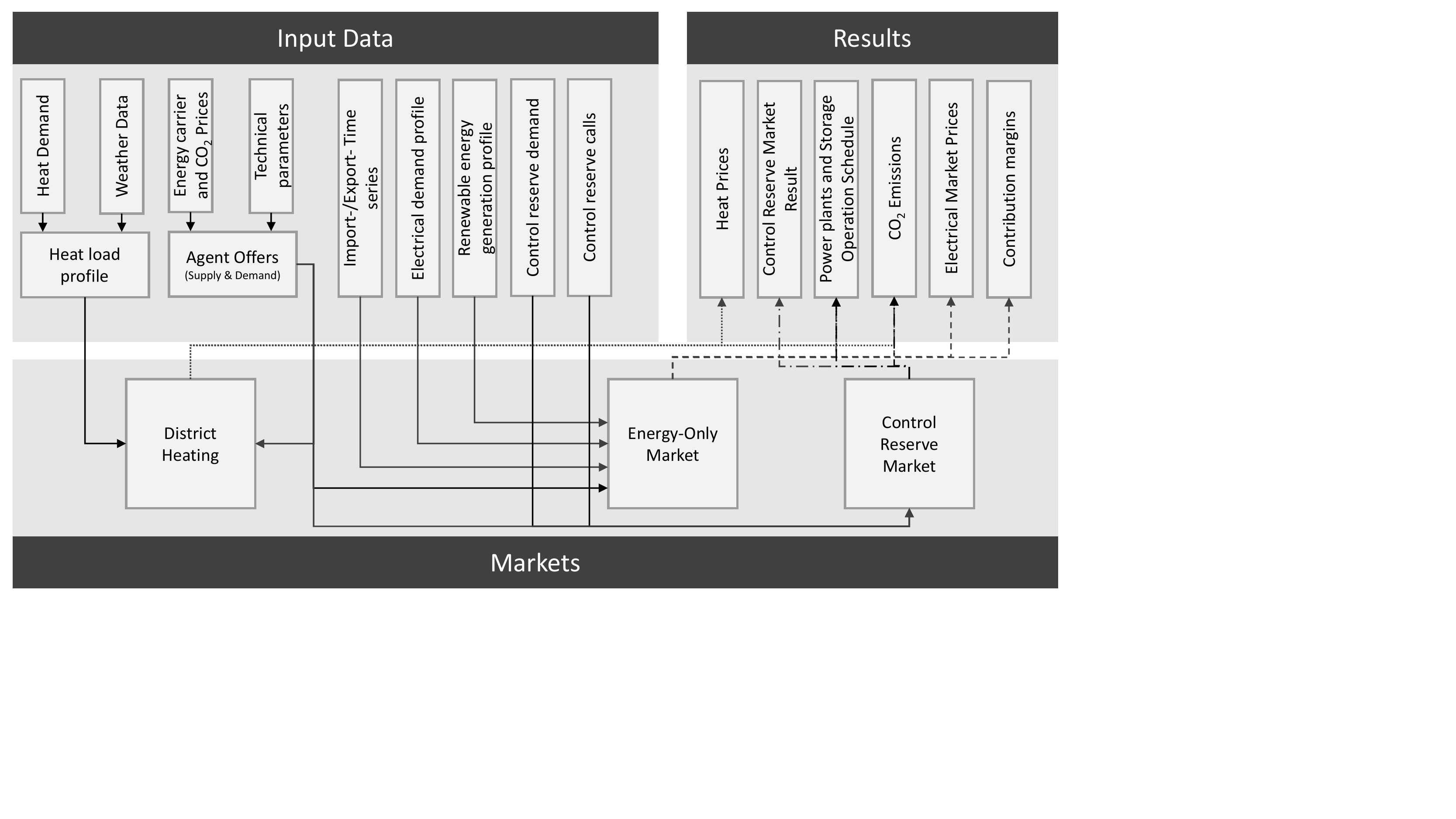}}
\caption{Overview of the structure of the market model}
\label{fig_ModelOverview}
\end{figure*}

The market model has a quarter-hour time resolution. It models an energy-only market coupled with a simplified control reserve procurement market. The two represent central electricity markets which operate independently from each other, and on which a large number of market players can act and develop their bidding strategies. The model also includes a heating market, where district heating is provided cost-efficiently to connected consumers. The model focus is on the German markets, while cross-border electricity transfers are accounted for through exogenous time series. The actors consist of fossil fuel-fired power plants and energy storage facilities. The supply from run-of-river, wind and photovoltaic power plants, which depends on meteorological factors, does not actively participate in trading in the chosen methodology. Instead, their feed-in is deducted in advance from the price-inelastic demand (exogenous load profile), which is equivalent to priority feed-in. A model structure overview is represented in Figure~\ref{fig_ModelOverview}, which shows the model exogenous input data, and the data flow in the model.

\subsection{Sequence of Actions}\label{subsec:sequence}
One simulation run of an entire year comprises several sequential steps. In a preparation phase, a simplified merit order model is calculated on the basis of nominal capacity and marginal costs of the fossil fuel-fired power plants and the residual load time series. This calculation is done for the purpose of generating a price forward curve for the entire year in 15 min time resolution. This price expectation is then available to all market participants for formulating their bids. In each time step (15 min) of the main phase, the markets are cleared in a defined order:
\begin{enumerate}
\item Demand from district heating is calculated, and cogeneration plants compete for serving the thermal load. 
\item Market participants that are technically able formulate their bids on the control reserve market, based on a defined bidding strategy. After clearing, they receive the information on the market result. 
\item Trading on the energy-only market takes place. Market participants determine their bids, again based on a defined bidding strategy. After clearing, they receive the information on the market result. 
\item The market results determine the operation of all plants, which form the starting point for the next iteration.
\end{enumerate}

In each time step, the current operational state of each plant, and the technical constraints -- for example ramping constraints -- form the basis of the bids. The plants' success on either of the markets determines the bidding strategy for the following market(s), and, finally, the operational state of the plant in the following time step. For example, a plant that successfully sold control reserve capacity needs to operate at a certain generation level, and therefore strives to sell the related output on the energy-only market. Following \cite{Weidlich2018}, market participants distinguish two parts of the capacity when formulating their bids: (i) the minimum capacity level which is necessary for avoiding cost-intensive shut-down and later restart, or for fulfilling obligations from previous markets (e.~g. the minimum stable output, if the plant has successfully sold positive control reserve), and (ii) the flexible capacity which can be supplied if prices are high enough to make the operation profitable. The corresponding bid price is determined as in \cite{Weidlich2018}. The bid price formulation takes into account opportunity costs (based on the price forward curve), possible restart cost (for the minimum capacity level) and variable generation cost (for the flexible capacity).

 Following this, an internal optimization of the power plant operation is done by each operator of a power plant portfolio, based on the market success. Control power deployment is decided on the basis of an exogenous time series and on the energy price of the control reserve bid. Finally, each agent determines the final net operational power of each of his plants for the current 15 min trading interval, before the same sequence is repeated for the next interval.

\section{Scenarios and Assumptions}\label{sec:scenarioOverview}

Three different coal phase-out scenarios were defined and simulated in the current analysis, which are summarized in Table~\ref{tab_Scenarios_BMU}. The scenarios also help to illustrate the model's potential regarding the conclusions that can be drawn from the market results. The year 2017 was used as a reference year for all scenarios. The different assumptions made are based on the final report of the Commission on Growth, Structural Change, and Employment assembled by the  German government \cite{BMWI2019} (in the following named Coal Commission). In the final report, a proposal for a complete coal phase-out by the year 2038 is given.

\begin{table}[h!]
\centering
\caption{Overview of the simulated scenarios}
\label{tab_Scenarios_BMU}
\begin{tabular}{lrrr}
\toprule
\multicolumn{1}{c}{} & Scenario S1 & Scenario S2 & Scenario S3 \\
\midrule
\multicolumn{4}{@{}l}{Fossil-fired generation capacity (difference to the reference 2017)}   \\
\midrule
Lignite & $-4.8$\,GW & $-10.9$\,GW & $-10.9$\,GW \\
Hard Coal & $-7.6$\,GW & $-14.9$\,GW & $-14.9$\,GW \\
Nuclear & no change & $-11.4$\,GW & $-11.4$\,GW \\
Natural Gas & no change & no change & $+20.0$\,GW \\
\midrule
\multicolumn{1}{@{}l}{Renewable capacity} & $91.9$\,GW & $187.5$\,GW & $187.5$\,GW \\
\midrule
\multicolumn{1}{@{}l}{Electricity demand} & $493.3$\,TWh/a & $493.3$\,TWh/a & $493.3$\,TWh/a \\
 \midrule
\multicolumn{1}{@{}l}{Fuel prices} & Ref. 2017 & Ref. 2017 & Ref. 2017 \\
 \bottomrule
\multicolumn{4}{l}{\textsuperscript{1} According to ENTSOE \cite{ENTSOE.2019}.}\\
 \multicolumn{4}{l}{\textsuperscript{2} Based on \cite{Graichen.2018}, applied to reference year 2017.}\\
\end{tabular}
\end{table}


All three alternative scenarios include a substantial reduction in installed capacity from coal-fired power plants. Since the shut-down of coal-based power plants is motivated by climate policy goals, it is assumed that older, comparatively less efficient power plants will be shut down first. The decision on shut-down in the scenarios is done on the basis of the plant's specific CO$_2$ emissions during operation. The total capacity decommissioned is shown in Table~\ref{tab_Scenarios_BMU}. Since only entire power plant units can be shut down, the decommissioned capacity in the model deviates slightly from the specifications of the Coal Commission in some cases.

Scenario S1 considers decommissioning according to the measures proposed by the Coal Commission until the end of 2022, where 4.8\,GW of lignite and 7.6\,GW should be shut down. Scenarios S2 and S3 also implement the measures proposed by the Coal Commission until the end of 2030, resulting in a total of 10.9\,GW lignite and 14.9\,GW hard coal stopping their operation. Also, both scenarios S2 and S3 take into account the implementation of the nuclear phase-out decided in 2011, by which the currently remaining nuclear power plant capacity of about 11.4\,GW will also be decommissioned.

In scenario S3, it is assumed that an expansion of natural gas-fired power plants happens, in contrast to scenario S2. The installed capacity of gas power plants is required to compensate for the capacity bottlenecks in scenario S2 and, thus, ensure the security of supply. The amount of additional capacity is based on the maximum shortfall of the energy-only market in scenario S2. This assumption also follows the request of the Coal Commission in its final report, i.~e. that investment incentives should be set in a timely manner so that there is no time divergence between the demand for secure generation capacity and the completion of sufficient new plants (cf. \cite[p.67]{BMWI2019}). Renewable energy feed-in from wind and photovoltaic plants remains unchanged in scenario S1, whereas scenarios S2 and S3 assumes a generation capacity expansion. The expansion followed in the definition of the scenarios is based on path proposed by \cite{Graichen.2018}.

All further input data of the validated reference year 2017 are equal across all scenarios. This implies that no changes in fuel prices, demand for electricity as well as output and capacity of energy storage facilities are assumed. This assumption ensures that the analysis of the effects on market dynamics is better comparable, and is not affected by further assumptions on the future. Also, it reduces the number of changed parameters, which makes it easier to trace causes and effects.

\section{Results and Discussion}\label{sec:results}
One pivotal objective of electricity trading is to ensure the security of supply, i.~e. the balance between production and consumption. The reduction in installed capacity and the development of variable renewable energy resources may lead to situations in which this balance can no longer be easily maintained. To make the gap in security of supply visible in the three alternative scenarios, the market model was used to quantify the number of time intervals (15 min) in which the market demand was not fulfilled, or in which the residual load is negative. This can be taken as an indication of how much additional controllable generation (or other forms of flexibility) is required when phasing out coal. 

The time intervals of unfulfilled market demand (deficit) are those quarters of an hour, in which demand could not be fully served, despite the fact that the overall installed generation capacity is higher than the peak load. If a deficit occurs, it is due to technical constraints that make it impossible for a plant operator to offer (more) electricity to the market. These constraints include maximum ramp rates and minimum shutdown time of the different power plants. If no other capacity was built than assumed in the respective scenario, this deficit would have to be covered by imports. Negative residual load refers to a situation in which generation from VRE exceeds the market demand, and the extra amount should be either curtailed or exported. 

The summaries of some simulation results are displayed in Table~\ref{tab_results_1}. It shows that while in scenario S1, there are only 25 time intervals with deficit in the market, with a total volume of about 4.9\,GWh, the considerable reduction in coal output along with the nuclear phase-out in scenario S2 has higher impacts on the security of supply. In nearly 2\,000 quarters of an hour, the demand on the market cannot be met from domestic generation, resulting in a deficit of about 2.8\,TWh/a (equivalent to 0.5\,\% of the annual demand). The maximum power deficit is quite substantial, with 19.6\,GW. However, if just this capacity (rounded to 20\,GW) is added to the system in the form of natural gas power plants, as it is assumed in Scenario S3, 69 time intervals with deficit remain, as the plants cannot all be available when needed due to their technical constraints. The result may not be very surprising, given that S2 and S3 assume the shut-down of 37.2\,GW of conventional generation in total, but is shows that the additionally installed 95.6\,GW of renewable capacity do not substantially decrease the need for controllable power plants -- or other forms of positive flexibility -- and that these are only deployed at a few times in a year.

\begin{table}[h!]
\caption{Key figures for deficits and negative residual loads on the energy-only market in scenarios S1-S3}
\label{tab_results_1}
\centering
\begin{tabular}{lrrr}
\toprule
\multicolumn{1}{c}{} & Scenario S1 & Scenario S2 & Scenario S3 \\
\midrule
\multicolumn{2}{@{}l}{Deficit} &  &  \\
\midrule
Time intervals (15 min) & $25$ & $1\,908$ & $69$ \\
Maximum power & $3.1$\,GW & $19.6$\,GW & $4.8$\,GW \\
Total energy & $4.9$\,GWh/a & $2\,836$\,GWh/a & $23$\,GWh/a \\ \midrule
\multicolumn{2}{@{}l}{Negative residual load} &  &  \\
\midrule
Time intervals (15 min) & $0$ & $3\,510$ & $3\,528$ \\
Maximum power & $0$\,GW & $42.9$\,GW & $42.9$\,GW \\
Total energy & $0$\,GWh/a & $7\,929$\,GWh/a & $7\,929$\,GWh/a\\
\bottomrule
\end{tabular}
\end{table}

\begin{table}[h!]
\centering
\caption{Mean base, peak and off-peak electricity prices for scenarios S1-S3 and reference case}
\label{tab_results_2}
\begin{tabular}{@{}lrrrr@{}}
\toprule
\multicolumn{1}{c}{} & Reference & S1 & S2 & S3 \\
\midrule
\begin{tabular}[c]{@{}l@{}}Base (EUR/MWh)\end{tabular} & $35.45$ & $47.70$ & $52.65$ & $45.11$ \\
\begin{tabular}[c]{@{}l@{}}Peak (EUR/MWh)\end{tabular} & $37.72$ & $53.06$ & $54.00$ & $47.17$ \\
\begin{tabular}[c]{@{}l@{}}Off-Peak (EUR/MWh)\end{tabular} & $33.07$ & $40.50$ & $51.57$ & $42.60$\\
\bottomrule
\end{tabular}
\end{table}

A comparison of resulting electricity market prices with the reference scenario shows that the reduction of coal-fired power generation increases prices in all three alternative scenarios. Table~\ref{tab_results_2} shows the summarized results for average (mean base), peak (workdays 8 a.m. -- 6 p.m.) and off-peak (remaining times) electricity prices. The highest increase in prices was seen in scenario S2, due to the sharp decline installed capacity of fossil-fired generation. Many situations were observed in which expensive (gas-fired) peak-load power plants are in merit, as cheaper large-scale power plant technologies are no longer available. These more expensive time intervals more than compensate the low market prices that occur when renewable supply is high, resulting in an overall price increase.

\begin{figure*}[ht]
\centering
\subfloat{\includegraphics[width=\linewidth]{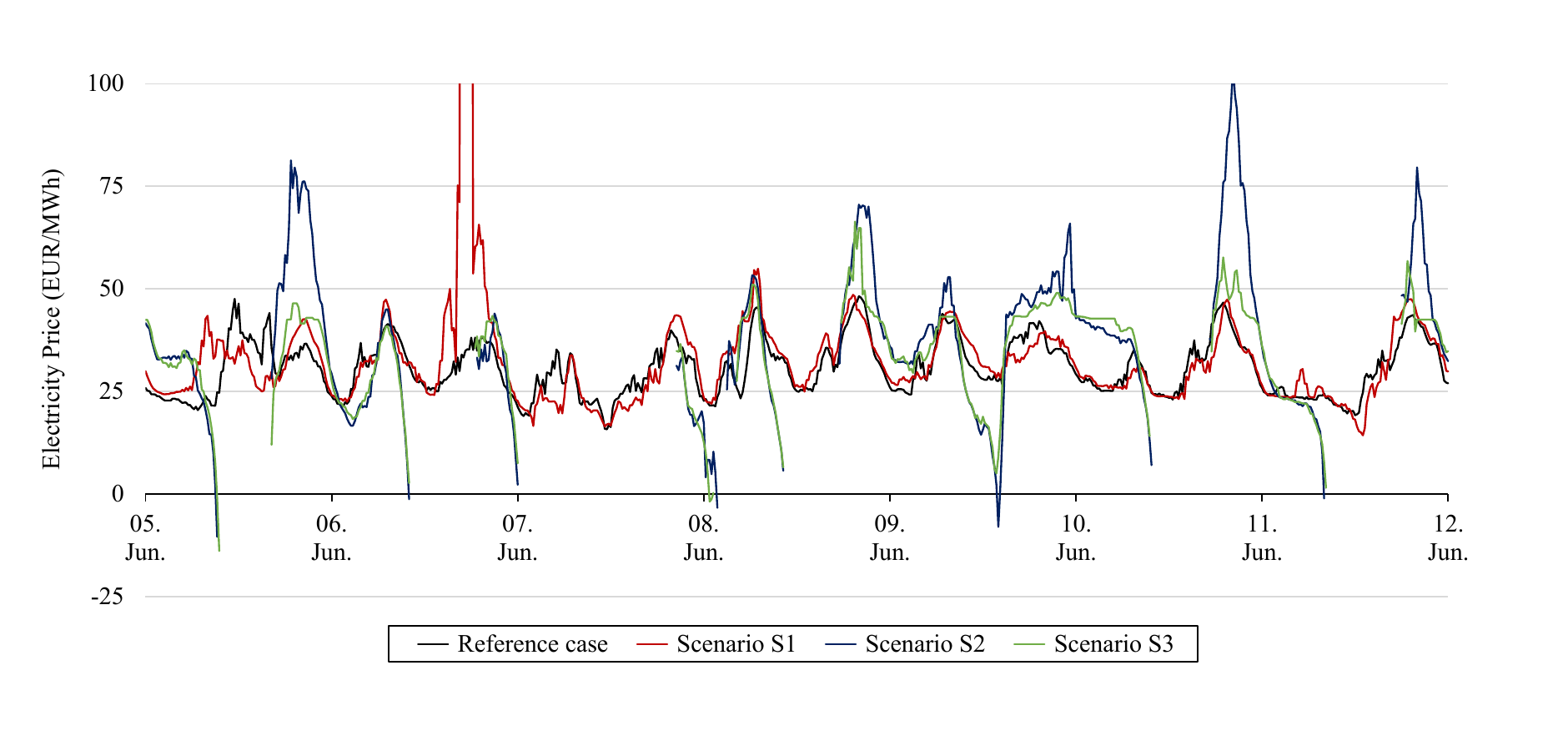}}
\caption{Detailed view of electricity prices in reference scenario, and scenarios S1 - S3, for a week in June with very high feed-in of renewable energy resources.}
\label{fig_PriceVolatility}
\end{figure*}

It can also be noticed that higher shares of VRE resources increase the electricity price volatility, as shown in Figure~\ref{fig_PriceVolatility}. The figure represents a comparison of the prices using an example week (June).  The effect of VRE share on price volatility is most strongly illustrated in scenario S2 since it is an extreme scenario (high VRE and low fossil-fired capacity). Discontinuities in the figure are due to time points with negative residual load. As VRE provide their power to the market at a price-independent bid, the lowest allowable price occurs in these intervals (here implemented as $-3\,000$\,EUR/MWh).

\begin{figure*}[hb]
\centering
\subfloat{\includegraphics[width=\linewidth]{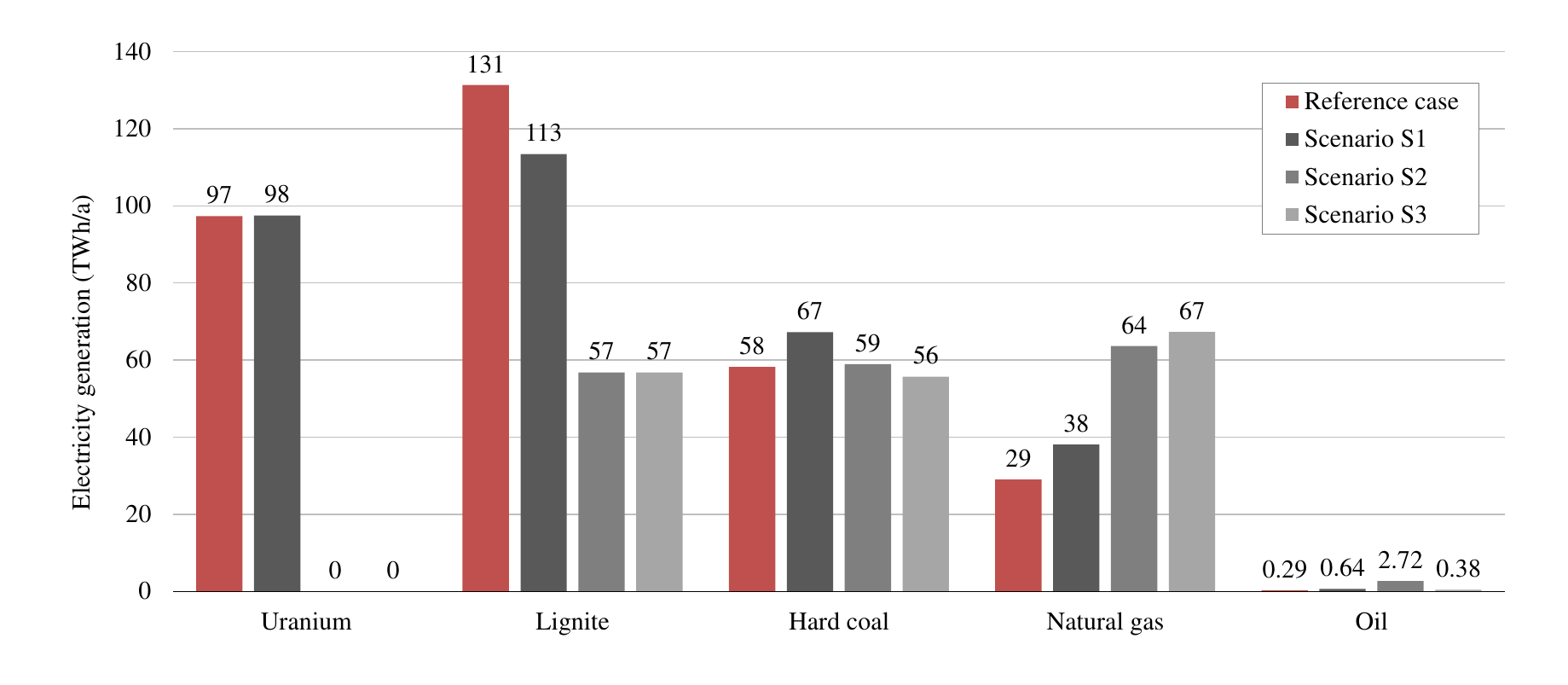}}
\caption{Energy source use of fossil-thermal power plants in scenarios S1 - S3 compared to reference data in TWh/a}
\label{fig_generationPerType}
\end{figure*}

The amount of energy generated by each technology varies between the scenarios. Figure~\ref{fig_generationPerType} shows that the usage of lignite and hard coal decreases in S1-S3 compared to the reference, in compliance with the goals of a coal phase-out. The only exception is the increase in electricity generation from hard coal power plants in scenario S1: despite a reduction of installed capacity by 7.6\,GW, electricity generation increases. This is due to the simultaneous decrease in the installed lignite capacity by 4.8\,GW, which moves the comparatively more expensive hard coal-fired power plants left in the merit order, thus increasing their frequency of market success.

Interestingly, the addition of 20\,GW of new natural gas-fired generation units in scenario S3 does not lead to a significant reduction in electricity generation from the remaining coal-fired power plants, due to the lower variable generation cost of the latter. Power generation from lignite is the same in S2 and S3, while generation from hard coal is only down by 3\,TWh ($\approx 5\,\%$) in S3 in comparison with S2. It can also be seen that the addition of gas power plants in scenario S3 does not lead to a significant increase in electricity generation from this energy source, from which a low frequency of use can be derived. The full load hours of the entire gas-fired portfolio fall from 2\,900\,h/a (S2) to 1\,600\,h/a (S3). This finding emphasizes that investments into new generation capacity might not happen without subsidies or some form of capacity mechanism, as not all plants are economically viable.

\section{Conclusion}\label{sec:conclude}
The results from a detailed electricity simulation model introduced here have shown that the currently discussed phase-out of coal-fired power generation in association with the expansion of renewable energy deployment will change market dynamics considerably. This process must, therefore, be backed by some strategy of facilitating investments into alternative controllable generation capacities or other forms of flexibility. Based on the assumptions made, it was shown that the increasing volatility of the market will lead to supply deficits, if no additional controllable generation capacity is installed. Otherwise, larger amounts of imports must be used to cover the observed shortages of domestic supply in the scenarios simulated here. Besides, a growing number of time steps with generation surplus was observed with the expansion of VRE resources. Adding another 20\,GW of gas-fired power generation capacity did not completely overcome the challenges, as shown in the scenario S3 defined in this work. 

The findings illustrate the need for additional flexibility options to be made available to the system for backing up the coal phase-out. It was also shown that these new investments might not all be profitable, as some capacity is only deployed in a few time intervals per year. This highlights the necessity of providing incentives for those investments that are necessary to ensure the security of supply in the future low-carbon German electricity system.

%

 \bibliographystyle{IEEEtran}
\bibliography{IEEEabrv,references}

\end{document}